\documentclass[pra,twocolumn,showpacs,amsmath,amssymb,superscriptaddress,unsortedaddress,floatfix]{revtex4}

\usepackage{graphicx,amsfonts,amsmath}% Include figure files
\usepackage{dcolumn}% Align table columns on decimal point
\usepackage{bm}% bold math
\usepackage{color}
\newcommand\ket[1]{|#1 \rangle}

\newcommand\ketbra[2]{|#1 \rangle\!\langle#2|}

\begin{document}

\preprint{APS/PRB}

\title{Decoherence enhanced quantum measurement
of a quantum dot spin qubit}% Force line breaks with \\

\author{Katarzyna Roszak}%
\affiliation{%
Institute of Physics, Wroc{\l}aw University of Technology,
50-370 Wroc{\l}aw, Poland
}%
\author{{\L}ukasz Marcinowski}
\affiliation{%
Institute of Physics, Wroc{\l}aw University of Technology,
50-370 Wroc{\l}aw, Poland
}%
\author{Pawe{\l} Machnikowski}%
\affiliation{%
Institute of Physics, Wroc{\l}aw University of Technology,
50-370 Wroc{\l}aw, Poland
}%

%\keywords{Suggested 
\begin{abstract}
We study the effect of phonons on a proposed scheme for the direct measurement
of two-electron spin states in a double quantum dot by monitoring the 
the noise of the current flowing through a quantum point contact coupled to 
one of the dots. We show that although the effect of phonons is damaging to the procedure at extremely low temperatures characteristic
of spin-in-quantum-dots experiments, and may
even be fatal, increasing the temperature leads
to a revival of the schemes usefulness. Furthermore, at higher, but still 
reasonably low temperatures
phonon effects become advantageous to the measurement scheme, and lead to the
enhancement of the spin-singlet noise without disturbing the low spin-triplet noise.
Hence, the uncontrollable interaction of the measured system with 
the open bosonic environment, can be harnessed to increase the distinguishability
between the measured states.
\end{abstract}

\pacs{73.21.La, 72.25.Rb, 63.20.kd, 03.67.Lx}% PACS, the Physics and Astronomy
                             % Classification Scheme.
%\keywords{Suggested 
\maketitle

Measurement processes have attracted a lot of attention recently, due to the 
growing need to stretch the attainable measurement precision, as well as 
the necessity to measure system properties which are hard to access.
This led to the development of a new field in quantum information science,
namely, quantum metrology \cite{giovannetti06,giovannetti11}, the purpose of which is to enhance measurement
capabilities by making the measurement device quantum itself. For the moment,
quantum metrology requires pure states, and the proposed schemes are
expected to react very badly to decoherence \cite{shaji,huelga,auzinsh,demkowicz,kolodynski}. 
It has been shown that the most popular entanglement
based scheme of quantum enhanced metrology suffers from an
unavoidable drawback due to quantum noise  which
reduces (for long chains of quantum systems)
the quantum (Heisenberg) metrology limit to the classical shot-noise
limit \cite{demkowicz12}.
Since in solid state systems
decoherence is unavoidable, a different approach is needed. It turns out that
in some situations, decoherence can be helpful (e.g. to facilitate electron
transport in quantum dot (QD) chains \cite{rebentrost,plenio08}), and may even be utilized 
for measurement purposes \cite{braun10,mazurek13}. Here, we discuss a system for which
a strong interaction with the environment
can lead to the enhancement of the distinguishability between the measured states,
and be beneficial for the measurement.

The system to be measured consists of two QDs, each containing an electron.
If the electrons are in a spin-triplet configuration, electron transitions 
between the dots are forbidden due to the Pauli principle, unless they can occupy
one of the excited QD levels. Since QDs are zero dimensional structures, all QD
realizations are very small with dimensions varying between a few and a few tens of 
nanometers in each direction. It follows that the first excited state of a QD is 
typically very distant energetically from its ground state. The energy difference
ranges from 10 meV in larger QDs to a 100 meV in small dots, hence, considering
the other energy scales characteristic for a double QD (DQD) system, excited QD states can be 
neglected in almost all situations.

In the case, when the DQD electrons are in a spin-singlet configuration,
electron transitions between the dots are allowed, and both electrons can
occupy a single dot. This is hindered by the Coulomb interaction between
the electrons which is much larger when both electrons occupy a single dot
than for singly occupied dots.
This repulsion is of the order of 1 meV, and the energy required to enable such a process
can be supplied, e.g., by unavoidable interactions with the solid state environments
of the QD. Hence, depending on the two-electron spin configuration, the occupation
of the QDs will either be constant (for the spin-triplet), or will fluctuate between
zero, one and two electrons localized in each dot (for the spin-singlet).

This difference in charge fluctuations 
is the basis for the direct quantum point contact (QPC) 
measurement of the two-electron
spin configurations which was proposed in Ref.~[\onlinecite{barrett06}].
The idea is, that since the current flowing through a QPC located near a QD
depends on the occupation of the QD via the Coulomb interaction between electrons
in the QD and the electrons in the QPC, the current fluctuations will be strongly affected
by the fluctuations of the QD occupation. Since triplet occupation cannot fluctuate,
the current flowing through a QPC coupled to one of the dots forming the DQD is Poissonian, 
as expected. Yet, if the DQD is in a singlet configuration, 
the interaction between the DQD and the QPC can induce inter-dot transitions and the
resulting QD occupation fluctuations in the coupled QD lead
to QPC current fluctuations and subsequently the current noise becomes strongly 
super-Poissonian. Therefore, by measuring the QPC
current noise one is able to distinguish between the spin-triplet and spin-singlet states.

The weakness of this measurement scheme lies in the fact that the QDs are embedded
in a solid state structure and the QD electrons are subject to an unavoidable interaction
with vibrations of the crystal lattice surrounding the dots (phonons) 
\cite{muller12,ramsay10,vagov04}. 
This interaction can also induce transitions between singly and doubly occupied 
QD states while it conserves two-electron spin symmetry \cite{grodecka08a}
(in fact, it causes exactly the same transitions as the DQD-QPC interaction \cite{roszak09}).
As shown in
Ref.~[\onlinecite{marcinowski13}], at very low temperatures which are characteristic
for experiments performed on electron-spin states confined in QDs (strongly sub-Kelvin)
\cite{averin05a,meunier06,rogge05,barthel09,cassidy07,bylander05,elzerman04} the electron-phonon interaction counteracts transitions to the high-energy
doubly occupied states.  This hinders the distinguishability of the measured states,
which relies on charge fluctuations. In the extreme case, when the electron-phonon
interaction is strong enough relative to the DQD-QPC coupling strength, 
the electron-phonon interaction can lead to the situation
when the singlet state is primarily composed only of the low energy configuration 
(with one electron in each dot), which cannot be distinguished from the triplet state
by the QPC.

In this paper we analyse the measurement scenario and the interaction with phonons
further and find, that the damaging effect can be overcome by increasing the 
temperature. Furthermore, for sufficiently high temperatures the electron-phonon
interaction can lead to an enhancement of the singlet noise, increasing the singlet-triplet
distinguishability instead of damaging it. Thus, we demonstrate a temperature-driven
transition from the regime of decoherence-induced suppression to decoherence-assisted enhancement
of the measurement in this setup.

Let us first set the framework for the theoretical description of the system
(following Refs.~[\onlinecite{barrett06,marcinowski13}]) by defining its Hamiltonian
and finding the equations of motion for its density matrix, which will then allow us 
to analyze the QPC current noise.
We do not include any interactions that could induce singlet-triplet transitions. 
If the excited QD states are energetically unavailable, there is only one triplet configuration
and the evolution in the triplet sector is trivial. Hence, in the following the focus is 
laid on the singlet subspace of the Hilbert space.

The Hamiltonian of the DQD system with two electrons is given by
\begin{displaymath}
H_{{\mathrm{DQD}}} = 
\Delta \sum_{\sigma=\uparrow , \downarrow}
(a_{R\sigma}^\dagger a_{L\sigma} + a_{L\sigma}^\dagger a_{R\sigma}) +
U \sum_{i =R,L} n_{i\downarrow} n_{i\uparrow},
\end{displaymath}
where $\Delta$ is the inter-dot tunnelling amplitude,
$a_{i\sigma}$, $a_{i\sigma}^\dagger$
are the annihilation and creation operators of an electron in dot $i = \mathrm{R,L}$ 
(right, left) with
spin $\sigma = $~$\uparrow,\downarrow$, $n_{i\sigma}=a_{i\sigma}^\dagger a_{i\sigma}$ 
gives the number of electrons with spin $\sigma$ in dot
$i$, and $U$ is the Coulomb charging energy for adding a second electron to a QD.
The eigenstates of this Hamiltonian are easily found and in the singlet subspace are 
given by
\begin{subequations}
\label{eqn:states}
\begin{eqnarray}
\ket{s_0} &=&  \xi' \Big(\ket{\uparrow \downarrow} -
\ket{\downarrow \uparrow}\Big) -
 \xi \Big(\ket{d_L} + \ket{d_R}\Big), \\ 
\ket{s_1} &=& \xi \Big(\ket{\uparrow \downarrow} -
\ket{\downarrow \uparrow}\Big)+ \xi' \Big(\ket{d_L} +
\ket{d_R}\Big),\\ 
\ket{s_2} & = & \frac{1}{\sqrt{2}} \Big(\ket{d_L} -
\ket{d_R}\Big),
\end{eqnarray}
\end{subequations}
where $\ket{\sigma \sigma'}= a^\dagger_{L\sigma}a^\dagger_{R\sigma'}
\ket{0}$ denote singly occupied states, and 
$\ket{d_{i}} = a^\dagger_{i\uparrow}a^\dagger_{i\downarrow}
\ket{0}$, $i=L,R$, are doubly occupied states. The parameters are equal to 
\begin{eqnarray}
\nonumber
\xi &=&\frac{1}{\sqrt{2}} \sin (\theta/2)\\
\nonumber
\xi' &=&
\frac{1}{\sqrt{2}} \cos (\theta/2), 
\end{eqnarray}
where $\theta
= \mathrm{atan}(4\Delta/U)$. 
The corresponding eigenenergies are equal to $-J, U, U+J$,
respectively,  where 
\begin{equation}
\nonumber
J = \frac{1}{2}(\sqrt{U^2 + 16
  \Delta^2} - U)
\end{equation} 
is the energy difference between the lowest 
energy singlet and the triplet states.
The triplet energy is
taken equal to zero, and serves as a reference for the calculated energies
of the singlet states.

The QPC is situated near one of the QDs in such a way that it is 
sensitive to the occupation of only this one (right) dot.
The DQD-QPC interaction is described by the Hamiltonian
\begin{equation}\label{equ:HamTun}
H_{{\mathrm{tun}}} = 
\sum_{p,q,\sigma} (T_{pq} + \chi_{pq}
n_R ) a_{Sp\sigma}^\dagger a_{Dq\sigma} + \mbox{H.c.},
\end{equation}
which accounts for the tunnelling of electrons through the QPC and contains
a factor dependent on the occupation of the right dot, $n_R$.
Hence, electron tunnelling consists of a robust part, independent of the QD occupation
described by the constants $T_{pq}$, and a Coulomb interaction induced enhancement $\chi_{pq}$. 
The tunnelling constants are assumed to be slowly varying over
the energy range where tunnelling is allowed \cite{gurvitz96,barrett06}
and are taken constant.
Here, $a_{np\sigma}$, $a_{np\sigma}^{\dagger}$ are the QPC electron
annihilation and creation operators corresponding to an electron in lead $n =
\mathrm{S,D}$ (source, drain) 
and in mode $p$, with the distinction
of spin $\sigma$ which is constant throughout the tunnelling.
In this paper, we study a QPC which operates in the
high bias regime, that is, in the situation when the chemical potential offset between the
leads is large enough to induce transitions to doubly excited states
\cite{barrett06}. 

The electron-phonon interaction Hamiltonian
is given by
\begin{equation}\label{equ:HamiltEPH}
H_{{\mathrm{e-ph}}} = \sum_{\sigma,i} \sum_{\bm{k} ,\lambda}
F_i^{(\lambda)} (\bm{k}) a_{i\sigma}^{\dagger} a_{i\sigma}
(b_{\bm{k},\lambda} + b_{-\bm{k},\lambda}^\dagger),
\end{equation}
where
$b_{\bm{k},\lambda}$ and $b_{\bm{k},\lambda}^{\dagger}$
are phonon annihilation and creation operators for a phonon from
branch $\lambda$ with wave vector $\bm{k}$.
$F_{\mathrm{L/R}}^{(\lambda)} (\bm{k}) = F^{(\lambda)}(\bm{k})
e^{\pm ik_x d/2}$
are electron-phonon coupling constants, and $d$ is the inter-dot distance. 
The coupling constants depend on material parameters, the types of phonons 
and electron-phonon couplings 
taken into account, and the electron wavefunction. Their explicit form can be found
in Ref.~[\onlinecite{marcinowski13}].

Following Refs.~[\onlinecite{barrett06,marcinowski13}], we find the quantum
master equation in the Lindblad form for the DQD interacting with 
both environments (the bosonic phonon environment and the fermionic
QPC environment) assuming that the environments are not correlated with each other, 
\begin{eqnarray}\label{equ:lindbladequation}
  \dot \rho (t)& = &\mathcal{L}(\rho)=- \frac{i}{\hbar} [H_{{\mathrm{DQD}}},\rho] 
\\ \nonumber & &+ 
\frac{1}{\hbar^2} \left (\sum_i^3 \frac{1}{2}(C_i^\dagger C_i \rho + \rho
C_i^\dagger C_i ) +  \sum_i^3 C_i\rho C_i^\dagger \right)
 \\  & &+ \frac{1}{\hbar^2} \left(\sum_i^4 \frac{1}{2}(B_i^\dagger B_i \rho + \rho
 B_i^\dagger B_i ) + \sum_i^4 B_i\rho B_i^\dagger \right).\nonumber
\end{eqnarray}
The Lindblad operators corresponding to the
DQD-QPC interaction are given by \cite{barrett06}
\begin{eqnarray*}
  C_1 & = & \nu \sqrt{\frac{V-(U+J)}{\hbar}} \sin\frac{\theta}{2} \ketbra{s_2}{s_0}, \\
  C_2 & = & \nu \sqrt{\frac{V+(U+J)}{\hbar}} \sin\frac{\theta}{2} \ketbra{s_0}{s_2}, \\
  C_3 & = & \sqrt{\frac{V}{\hbar}} \left[ (\mathcal{T} + \nu)\mathbb{I} +\nu \cos\frac{\theta}{2} 
\left(\ketbra{s_1}{s_2} + \ketbra{s_2}{s_1}\right) \right],
\end{eqnarray*}
where $V = (\mu_S - \mu_D)$ is the QPC bias, 
$\mathcal{T}=\sqrt{4\pi g_Sg_D}T$ is the unconditional tunnelling constant,
with $T=T_{pq}$, and $\nu=\sqrt{4\pi
  g_Sg_D}\chi$ is the 
tunnelling constant conditioned on the occupation of the right QD,
with $\chi=\chi_{pq}$. Here, $g_i$ is the density of states of the $i$-th QPC lead ($i=S,D$).

The Lindblad operators for the electron-phonon interaction 
are of the standard form, $B_n = \sqrt{\gamma_{ij}} \ketbra{s_i}{s_j}$.
Two operators correspond to transitions between $|s_0\rangle$ and $|s_1\rangle$
when $i=0$, $j=1$ and $i=1$, $j=0$,
while the other two describe transitions between $|s_1\rangle$ and $|s_2\rangle$
when $i=1$, $j=2$ and $i=2$, $j=1$. 
The transition rates $\gamma_{ij}$ are calculated using the Fermi golden rule
based on the Hamiltonian in eq. (\ref{equ:HamiltEPH}), with the usual temperature 
dependence $\gamma_{ij}(T)=\gamma_{ij}(T=0)[n(\omega_j-\omega_i,T)+1]$,
where $n(\omega,T)$ is the Bose distribution and
$\hbar\omega_i$ are the energies corresponding to
a given singlet state.

The time evolution of the DQD system is found by solving the above generalized
master equation. Specifically, the singlet steady state, found from
the condition $\dot \rho (t)=0$, is
\begin{eqnarray}
\label{ssss}
\rho_\infty^{(\mathrm{s})} &=& \frac{1}{N}
\left[(A_+^2 + \gamma_{02})(A_0^2 + \gamma_{21})\ketbra{s_0}{s_0}
  \right .\\ & &
+ (A_0^2 + \gamma_{12})(A_-^2 + \gamma_{20})\ketbra{s_1}{s_1}
\nonumber  \\ & &+
\left . (A_-^2 + \gamma_{20})(A_0^2 + \gamma_{21})\ketbra{s_2}{s_2}\right ],  \nonumber 
\end{eqnarray}
where 
\begin{eqnarray}
\nonumber
N &=& (A_+^2 + \gamma_{02})(A_0^2 +\gamma_{21}) + (A_0^2
+\gamma_{12})(A_-^2 + \gamma_{20}) \\
\nonumber
&&+(A_-^2 + \gamma_{20})(A_0^2 + \gamma_{21}),\\
\nonumber
A_{\pm} &=& \nu \sqrt{(V\pm J \pm U)/\hbar} \sin (\theta/2),\\
\nonumber
A_0 &=&\nu  \sqrt{V/\hbar} \cos (\theta/2).
\end{eqnarray}
The singlet steady state depends explicitly on the strengths of the phonon-couplings
relative to the QPC coupling strengths.

The interaction between the DQD and the QPC eventually leads to the localization
of any two-electron initial state in either the singlet or the triplet subspace
\cite{barrett06}. This process is not disturbed by the presence of phonons, and localization
occurs on timescales of tens of microseconds \cite{marcinowski13}. 
Since the excited QD states are not taken into account, the QPC current noise
for a triplet state
remains Poissonian regardless of temperature due to the Pauli exclusion principle,
and the triplet Fano factor is always equal to one (see 
Ref.~[\onlinecite{barrett06,marcinowski13}] for details). 
Hence, to quantify
the distinguishability of the singlet and triplet states, it is sufficient to
study the QPC current noise for the singlet steady state. To this end we will
examine the Fano factor for the current noise corresponding to state (\ref{ssss})
and compare it to unity, which is the Fano factor value for Poissonian noise,
characteristic for the triplet states.

The Fano factor is defined as \cite{fano47,blanter01}
$F={S(0)} / {2e\bar{I}}$, where $\bar{I}$ is the mean current.
The QPC noise power spectrum for steady states is given by 
\begin{equation}
\nonumber
S(\omega) = 2
\int_{\infty}^{\infty} \mathrm{d}\tau G(\tau) e^{-i\omega \tau}. 
\end{equation}
The
the current-current correlation function, given by
\begin{equation}
G(t,t') = \langle I(t') I(t)\rangle - \langle I(t')\rangle
 \langle I(t)\rangle,
\label{cc}
\end{equation}
depends only on $\tau=t'-t$ in the long-time limit.
Following Refs.~[\onlinecite{goan01,barrett06}] we can find the
correlation function
for the singlet steady state,
\begin{eqnarray}
\nonumber
G(\tau) & = &  e^2 \left[\mathrm{Tr} \{\sum_{i,j}
  C_{j} [e^{\mathcal{L}(\tau-\Delta t)} C_i\rho^{(s)}_{\infty}C_i^\dagger]
  C_{j}^\dagger\}\right. \\ 
\nonumber
& & - \mathrm{Tr} \{ \sum_i C_i \rho^{(s)}_\infty
  C_i^\dagger\}^2+ \left.\mathrm{Tr} \{ \sum_i C_i \rho^{(s)}_\infty
  C_i^\dagger\} \delta(\tau)\right]. 
\end{eqnarray}

The results presented in this paper are obtained using parameters corresponding to GaAs structures
\cite{barrett06,loss98}. 
The DQD energies are taken $U = 1$ meV and $J = 0.1$ meV, and the QPC bias is taken to be
$V = 2$ meV.
The
zero-temperature phonon transition rates are $\gamma_{02}(T=0) = 1.15 \cdot
10^{-3}$ ns$^{-1}$ and
$\gamma_{21}(T=0) = 6.01 \cdot 10^{-8}$ ns$^{-1}$.
The scaling parameter $\alpha= \frac{\mathcal{T}_0}{\mathcal{T}} =
\frac{\nu_0}{\nu}$, with $\mathcal{T}_0 = 0.1$ and $\nu_0 = 2.25 \cdot
10^{-3}$, is chosen in such way that $\alpha = 1$ corresponds to the
situation when the interaction with phonons is roughly the same
strength as the interaction with the QPC, 
meaning that $\sqrt{\gamma_{02}}=\nu_0 \sqrt{V/\hbar}$.
Then phonons dominate at large
$\alpha$ (small QPC current), while the DQD-QPC interaction dominates for 
small $\alpha$.

\begin{figure}[tb]
  \includegraphics[scale=0.7]{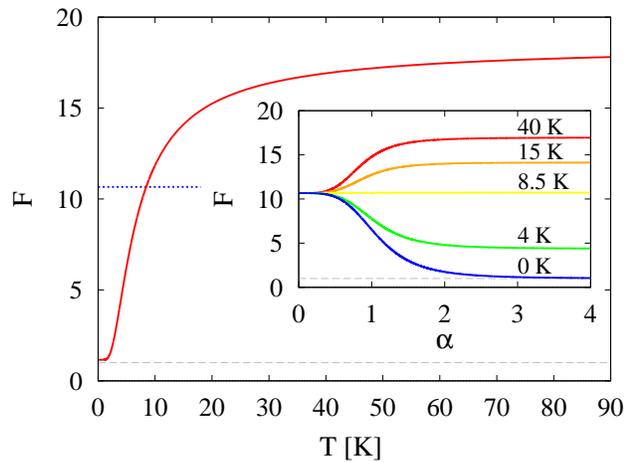}
  \caption{The Fano factor for a relatively strong electron phonon-interaction,
$\alpha=3$, as a function of temperature (red, solid line); no-phonon Fano factor
(blue, dotted line).
Inset: The Fano factor as a function of the strength of the electron-phonon
interaction relative to the DQD-QPC interaction, $\alpha$, for different
temperatures. In both plots, the dashed grey line is set at Fano factor equal to one
(Poissonian noise).}
  \label{fig}
\end{figure}

Fig. \ref{fig} shows the dependence of the singlet Fano factor as a function of temperature
for relatively strong electron-phonon interaction at $\alpha = 3$;
Fano factor equal to one corresponding to the triplet state is marked by the dashed
grey line. Although at $T=0$ phonons of such strength completely 
destroy the distinguishability of the
singlet and triplet state via QPC current noise (both singlet and triplet Fano factors
are equal to one), this is remedied already by a slight increase of the temperature.
At $T=8.5$ K, the redistribution of the different singlet occupations is strong enough to
restore the Fano factor to the no-phonon value found
(the no-phonon value of the Fano factor is marked by the blue, dotted line).
This is
due to the rising importance of phonon-induced transitions to higher energy, doubly
occupied states 
which compensate for the transitions to the lower energy, singly
occupied state. 
Further increasing the temperature leads to an enhancement of the Fano factor beyond 
the no-phonon value. The effect of temperature is strong between 0 and 20 K, and then 
starts to slowly saturate, yielding already ample enhancement of the Fano factor by 15 K,
long before the excited QD states need to be taken into account.

This is further quantified in the inset of the figure where the Fano factor is plotted
as a function of the scaling parameter $\alpha$, which is 
inversely proportional to the strength 
of the DQD-QPC interaction, at different temperatures. Regardless of the temperature, 
phonon effects are negligible when the DQD-QPC interaction dominates, near $\alpha=0$.
Their importance rises quickly, when the interactions become comparable, and then
saturates slowly after the interaction with phonons becomes twice as strong as
the interaction with the QPC. Hence, the interaction with phonons is relevant
when the QPC is coupled weakly to the DQD, and may serve to increase the 
effectiveness of the measurement in this situation.

At high enough temperatures, the electron-phonon interaction will 
cause transitions
to excited QD states.
This will lead to double occupations in the triplet subspace
in the situation when one of the electrons is in the QD ground state, and the other
is in the QD excited state. The occurrence of transitions to such states will lead to 
fluctuations of the QPC current noise, which will also become super-Poissonian.
Although at not extremely high temperatures this super-Poissonian character
will be much weaker than in the singlet case, it could complicate data interpretation.
However, since the dependence on temperature is strong only for relatively small temperatures,
the high temperature range 
is of no experimental interest.

We have studied the temperature dependence 
of a proposed scheme for the direct measurement of the DQD spin-singlet and spin-triplet 
states using the noise of the current flowing through a QPC coupled to one of the QDs.
We show, that although at low temperatures the interaction with phonons
has a destructive effect on the measurement scheme, the situation may be inverted
by a small increase of the temperature. Furthermore, beyond $T=8.5$ K
phonon-induced transitions actually help the distinguishibility of the singlet and
triplet states, leading to an enhancement rather than destruction of measurement
capability. 
Already at 15 K, the Fano factor, which is a measure of noise power 
relative to Poissonian noise power, 
can be raised by a factor of 1.32, and by a factor of 1.58
at 40 K, meaning that the effect occurs already at temperatures far below the limit when
the excited states of the QDs need to be taken into account.
Hence, we have shown that the unavoidable phonon-induced decoherence mechanism
can be used to facilitate measurement.

This work was supported in parts by the TEAM programme of 
the Foundation for Polish Science, co-financed from the European Regional 
Development Fund and by the Polish NCN Grant No. 2012/05/B/ST3/02875.

\end{document}